\documentclass[reprint,prb,twocolumn,showpacs,superscriptaddress,aps,longbibliography]{revtex4-1}
\usepackage[utf8]{inputenc}
\usepackage{graphicx}
\usepackage[colorlinks=true,citecolor=blue]{hyperref}
\usepackage{amsmath}
\usepackage{amssymb}
\usepackage{mathtools}
\usepackage{soul}
\usepackage{gensymb}
\usepackage{amsfonts,amssymb,amsbsy}
\usepackage{siunitx}
\usepackage{float}
\usepackage{physics}
\usepackage{booktabs}
\usepackage[percent]{overpic}

\newcommand{\dIdV}{d$I$/d$V$~}
\usepackage{xcolor}
\usepackage{natbib}

\usepackage{lineno}

\makeatletter
\def\@ziying@name{Ziying Wang}%
\def\doauthor#1#2#3{%
  \ignorespaces#1\unskip\@listcomma
  \begingroup
   #3%
  \@if@empty{#2}{\endgroup{}{}}{\endgroup{\comma@space}{}\frontmatter@footnote{#2}}%
  \def\@tempz{#1}%
  \ifx\@tempz\@ziying@name
   \textsuperscript{,\,}\frontmatter@footnotemark{2}%
  \fi
  \space \@listand
}%
\makeatother

\begin{document}

\title{Interaction-driven electronic ferroelectricity in van der Waals heterostructures}

\author{Ziying Wang}
\thanks{These authors contributed equally.}
\affiliation{Aalto University, Department of Applied Physics, 00076 Aalto, Finland}

\author{Ana Vera Montoto}%
\thanks{These authors contributed equally.}
\affiliation{Aalto University, Department of Applied Physics, 00076 Aalto, Finland}

\author{Mohammad Amini}
\affiliation{Aalto University, Department of Applied Physics, 00076 Aalto, Finland}

\author{Yuxiao Ding}
\affiliation{Aalto University, Department of Applied Physics, 00076 Aalto, Finland}

\author{Jose L. Lado}%
\affiliation{Aalto University, Department of Applied Physics, 00076 Aalto, Finland}

\author{Robert Drost}
\affiliation{Aalto University, Department of Applied Physics, 00076 Aalto, Finland}

\author{Adolfo O. Fumega}%
\email{Corresponding authors. Email: ziying.wang@aalto.fi, adolfo.oterofumega@aalto.fi, peter.liljeroth@aalto.fi}
\affiliation{Aalto University, Department of Applied Physics, 00076 Aalto, Finland}

\author{Peter Liljeroth}
\email{Corresponding authors. Email: ziying.wang@aalto.fi, adolfo.oterofumega@aalto.fi, peter.liljeroth@aalto.fi}
\affiliation{Aalto University, Department of Applied Physics, 00076 Aalto, Finland}

\begin{abstract}

Strong electronic correlations in narrow-band systems provide a promising route to realize emergent quantum phases. While ferroelectricity in van der Waals materials is typically associated with inversion symmetry breaking driven by lattice distortions, interlayer sliding, or moiré reconstruction, the possibility of generating ferroelectricity directly from electronic interactions remains largely unexplored. Here, using molecular beam epitaxy, scanning tunneling microscopy, and \emph{ab initio} calculations, we investigate two stacking geometries of bilayer 1T-TaSe$_2$, A-C and A-C$'$, formed by coupled Star-of-David charge density wave phases.
We show that both stackings realize quasi-one-dimensional interacting chains, but are governed by distinct interaction mechanisms. In the A-C stacking, strong interlayer hybridization leads to dimerization and the formation of a band insulating state. In contrast, the A-C$'$ stacking is dominated by interlayer Coulomb interactions, producing a spontaneous charge imbalance between layers that gives rise to an out-of-plane ferroelectric polarization. Furthermore, we demonstrate that ferroelectric and antiferroelectric interchain configurations can be stabilized and electrically switched by an external field. Our results prove that bilayer 1T-TaSe$_2$ is a platform for interaction-driven electronic ferroelectricity,
establishing an overlooked family of charge-ordered correlated states in 1T-TaSe$_2$ multilayers.

\end{abstract}

\maketitle

\section*{Introduction}

Ferroelectric materials, characterized by a spontaneous electric polarization, 
provide a rich platform for fundamental physics, and ultimately
a potential broad technological impact in nanoelectronics.\cite{Scott1989,Mathews1997}
The coupling between electronic, structural, and dielectric degrees of freedom makes them promising platforms for non-volatile memories, neuromorphic computing, sensors, and energy-efficient multifunctional devices.\cite{Garcia2009,Chanthbouala2011,Chanthbouala2012,Si2019,Khan2020,Wu2023,Martemucci2025}
The discovery of atomically thin ferroelectrics has extended these possibilities into the two-dimensional (2D) limit, enabling the integration of ferroelectric functionality into van der Waals (vdW) heterostructures.\cite{Zhang2022}
Several mechanisms have been shown to stabilize ferroelectricity in the monolayer limit, including phonon-driven displacive ferroelectrics,\cite{Chang2016,Yuan2019,Higashitarumizu2020,PhysRevLett.120.227601,Brehm2019,Abdelwahab2022,HuangFu2024} orbital-order ferroelectrics,\cite{Deng2025,Camerano2025} and magnetically driven ferroelectrics arising from the interplay between spin-orbit coupling and spin spirals.\cite{Song2022,Amini2024}
Beyond intrinsically polar 2D compounds, the tunability of vdW materials has enabled the engineering of ferroelectricity through interlayer coupling and stacking design.
In vdW heterostructures, ferroelectricity can emerge from polar stacking configurations or relative interlayer sliding that breaks inversion symmetry, giving rise to stacking and sliding ferroelectricity.\cite{Fei2018,Sharma2019,Zheng2020,Yasuda2021,ViznerStern2021,Akamatsu2021,Weston2022,Wang2022,Winterer2024}
More recently, moiré engineering has led to unconventional polar textures and emergent ferroelectric phases.\cite{SnchezSantolino2024,Anto2024,Li2025,Pan2025}
Despite this progress, most vdW ferroelectrics rely on structural inversion symmetry breaking associated with lattice distortions, interlayer sliding, or moiré reconstruction. In contrast, ferroelectricity driven directly by strong electronic correlations in flat-band systems remains elusive\cite{2025arXiv250815909D,2026NanoL26.4642S}.

Among vdW materials, the family of 1T transition metal dichalcogenides (TMDs) hosting Star-of-David (SoD)-type charge density wave (CDW) order provides a promising platform to explore flat-band states dominated by strong electronic correlations.\cite{Vao2021,Wan2023,Crippa2024,PhysRevLett.134.046504}
In these systems, the commensurate CDW reconstruction generates narrow electronic bands near the Fermi energy, driving a variety of correlated quantum phases.
In particular, monolayer 1T-TaSe$_2$ undergoes at low temperatures a commensurate CDW transition in which thirteen Ta atoms condense into a SoD superstructure.\cite{Nakata2021}
This reconstruction produces an isolated half-filled flat band with one unpaired electron per SoD cluster, leading to a Mott insulating ground state stabilized by electron-electron interactions.\cite{Chen2020}
The resulting triangular lattice of localized moments has further motivated proposals of exotic magnetic phases in the monolayer limit.\cite{Law2017,Ruan2021}
Interestingly, previous works have shown that interlayer stacking strongly modifies the electronic state in layered 1T compounds.
The relative stacking of neighboring SoD clusters controls the interlayer hybridization strength and can drive transitions between Mott insulating and band insulating regimes.\cite{ritschel2018stacking,butler2020mottness,Wang2020,PhysRevLett.129.016402}
More recently, studies have reported the emergence of moiré correlated states on twisted bilayers.\cite{Liu2026,AVMontoto2026}
Together, these results establish stacking engineering in bilayer 1T-TaSe$_2$ as a promising route to realize novel correlated phases, including interaction-driven ferroelectricity emerging from flat-band electronic states.

\begin{figure*}[ht!]
    \centering
    \includegraphics[width =\textwidth]{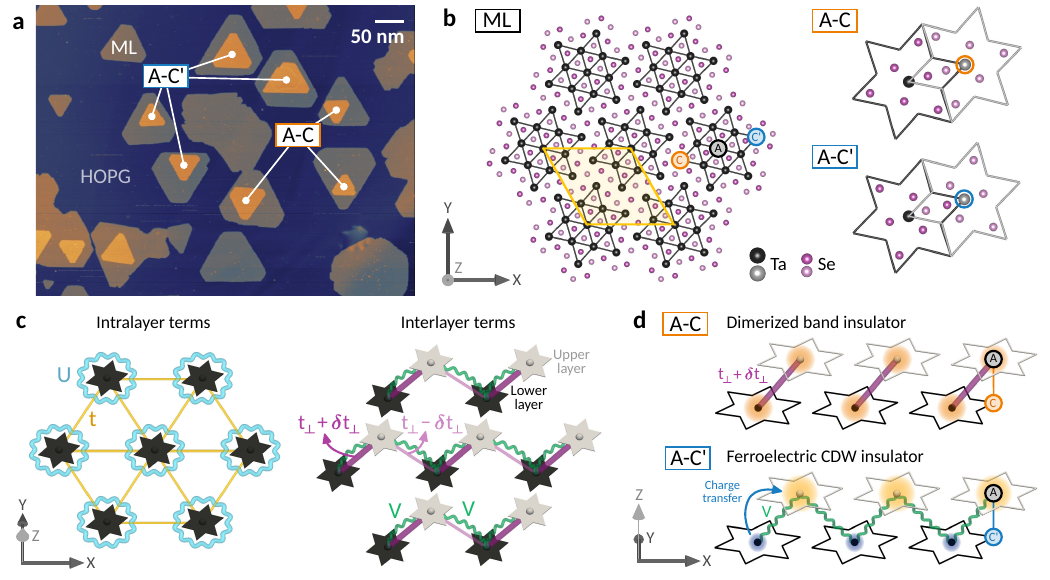}
    \caption{\textbf{a} Large-scale STM image of monolayer (ML) and different bilayer configurations (A-C and A-C$'$) of 1T-TaSe$_2$ on an HOPG sample. $V= 1$ V, $I = 10$ pA. \textbf{b} Top view of the SoD structure in monolayer 1T-TaSe$_2$ and the bilayer stacking arrangements A-C and A-C$'$. For clarity, only the Se atoms directly above the lower SoD (dark-colored) and below the upper SoD (light-colored) are depicted. \textbf{c} Schematic of the minimal tight-binding model for A-C and A-C$'$ bilayers as quasi-1D interacting fermionic chains, where each star represents a single orbital at the center of a SoD. The intralayer model's parameters include small nearest-neighbor hopping $t$ and on-site $U$ interaction. The interlayer terms are first and second nearest-neighbor hoppings  $t_{\perp} + \delta t_{\perp}$ and $t_{\perp} - \delta t_{\perp}$ and interlayer interactions $V$. \textbf{d} Schematic of the emergent phases in the different bilayer configurations. A dimerized band insulating phase occurs in the A-C bilayer due to its strong asymmetric interlayer hopping ($\delta t_{\perp} > 0$) and a ferroelectric CDW phase arises in the A-C$'$ bilayer due to the dominant interlayer interactions $V$. The charge transfer from one to the other layer is highlighted with blue and yellow spheres in the chain. This imbalance leads to an emergent ferroelectric polarization along the z direction.}
    \label{fig:fig1}  
\end{figure*}

In this work, we use molecular beam epitaxy (MBE) to grow bilayer 1T-TaSe$_2$ in two stacking geometries, A-C and A-C$'$.
Using scanning tunneling microscopy and spectroscopy (STM/STS) supported by \emph{ab initio} calculations, we show that the two configurations realize quasi-one-dimensional (1D) interacting fermionic chains.
The A-C stacking is dominated by interlayer dimerization, leading to a band insulating phase with a hybridization gap.
In contrast, the electronic order in the A-C$'$ stacking is governed by interlayer Coulomb interactions, producing a charge imbalance between layers that gives rise to an out-of-plane ferroelectric polarization.
Furthermore, we show that antiferroelectric and ferroelectric interchain configurations can be stabilized in the A-C$'$ phase and can be electrically switched by an external field.
Our work extends the phase diagram of bilayer SoD TMDs with a ferroelectric CDW phase driven purely by electronic interactions, providing a new bilayer platform for electrically tunable correlated states.

\section*{Results and Discussion}

\subsection*{Bilayer 1T-TaSe$_2$ as interacting quasi-1D fermionic chains}

\begin{figure*}[ht!]
    \centering
    \includegraphics[width = \textwidth]{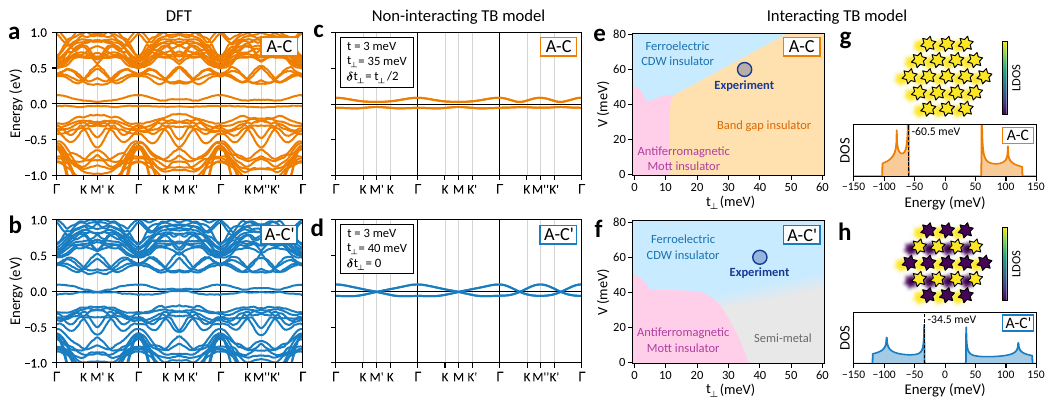}
    \caption{Simulated band structures and predicted phase diagrams and LDOS of the 1T-TaSe$_2$ bilayers from DFT and fitted tight-binding model.
    \textbf{a,} \textbf{b} DFT-simulated band structures of A-C  (a) and A-C$'$ (b) bilayers. \textbf{c,} \textbf{d,} Band structures of the A-C (c) and A-C$'$ (d) bilayers from the non-interacting tight-binding model with parameters fitted to the DFT results. \textbf{e,} \textbf{f} Phase diagrams of the A-C (e) and A-C$'$ (f) bilayers from the interacting tight-binding model, with on-site interactions $U = 100$ meV. The regions corresponding to the experimental results are highlighted. \textbf{g,} \textbf{h} Simulated DOS of the A-C (e) and A-C$'$ (f) bilayers from the interacting tight-binding model in the experimental region ($V = 60$ meV), and LDOS of each site at the energies marked with dotted lines. The lower-layer sites are blurred.}
    \label{fig:fig2}  
\end{figure*}

Figure \ref{fig:fig1}a shows the bilayer 1T-TaSe$_2$ sample grown by MBE on a highly oriented pyrolytic graphite (HOPG) substrate. In addition to monolayer (ML) islands, two bilayer stacking configurations are identified as A-C and A-C$'$ (see below for full experimental characterization of the structures). These bilayers present the same relative positions of Ta atoms, but with differing Se atomic positions, as illustrated in the structures shown in Fig.~\ref{fig:fig1}b.
Notably, these stacking configurations break the $C_3$ symmetry of the monolayer, effectively creating quasi-1D chains along the $x$-direction.
These structures result in strikingly different electronic behaviour, which can be described by a minimal tight-binding (TB) model for quasi-1D interacting chains, depicted schematically in Fig.~\ref{fig:fig1}c. It takes the form $H = H_0 + H_U + H_{V}$, where the single-particle hopping term $H_0$ accounts for the intralayer hopping between nearest neighbors $t$ and the hopping between interlayer first neighbors $t_{\perp} + \delta t_{\perp}$ and second neighbors $t_{\perp} - \delta t_{\perp}$. 
For the interacting part, we must take into account the onsite Coulomb interaction $U$ driving the Mott state in the monolayer, as well as the Coulomb interactions $V$ between the closest neighboring sites: the interlayer first- and second-nearest neighbors.

The resulting equations for the full Hamiltonian of the bilayer systems $H$ can be solved self-consistently by performing a mean-field approximation\cite{pyqula} (details in the methods section). In this way, and due to the relatively small in-plane hopping $t$ as compared to the interlayer hopping, this model describes a series of quasi-1D chains that link alternating upper and lower layer sites. As detailed later, the main difference between the two bilayer configurations is that we find significant dimerization in the A-C case ($\delta t_{\perp} = t_{\perp}/2$) leading to a dimerized band insulator, whereas the A-C$'$ bilayer is well captured with $\delta t_{\perp} = 0$. In this case, $V$ dominates, producing a charge transfer from one of the layers to the other, thus creating an emergent ferroelectric CDW phase with an out-of-plane polarization (Fig.~\ref{fig:fig1}d).

To get a more quantitative description of the different A-C and A-C$'$ stacking configurations,  we performed first-principles density functional theory (DFT) calculations (details in the methods section). Figures \ref{fig:fig2}a and b show the DFT band structures of both of these bilayers. The non-interacting part of the effective model $H_0$ can be matched to the corresponding DFT bands (Fig.~\ref{fig:fig2}c-d). With this method, we find good agreement for intralayer hopping $t = 3$ meV, much less dominant than the interlayer hopping parameters: $t_{\perp} = 35$ meV with broken symmetry $\delta t_{\perp} = t_{\perp}/2 = 17.5$ meV in the A-C bilayer, and $t_{\perp} = 40$ meV with no dimerization for A-C$'$.

Introducing the interacting terms characterized by $U = 100$ meV and $V$ into the tight-binding model, we extract the phases shown in Figure \ref{fig:fig2}e-f of the A-C and A-C$'$ 1T-TaSe$_2$ bilayers, for varying interlayer interaction $V$ and interlayer hopping $t_\perp$ (the mean-field results of the order parameters used to create these phase diagrams can be found in the SI Section A).
The A-C case considers some hopping anisotropy, $\delta t_{\perp} = t_{\perp}/2$, while the A-C$'$ case represents uniform hopping in the chains, $\delta t_{\perp} = 0$. For realistic values of $t_\perp$, in both cases, the antiferromagnetic Mott state present in the monolayer is suppressed. The dimerization of the chains in the A-C bilayer results in a band-insulating state, whereas the A-C$'$ bilayer is driven by $V$ from a semi-metallic state to a striped CDW phase.
Figure \ref{fig:fig2}g shows the total density of states (DOS) of the A-C bilayer predicted by the TB model in the band-insulating experimental regime ($t_{\perp} = 35$ meV, $V = 60$ meV), along with the local density of states (LDOS) map at energy $-60.5$ meV, which is uniform in all sites. The same DOS plot for the A-C$'$ bilayer in Figure \ref{fig:fig2}h ($t_{\perp} = 40$ meV, $V = 60$ meV) shows a smaller gap, and the LDOS map at energy $-34.5$ meV reveals the striped CDW phase. The stripe CDW phase corresponds to the quasi-1D chains of the TB model with charge transfer ($|\Delta n| \geq 0.97$) between the lower and upper layer sites. The charge transfer produces an out-of-plane polarization whose direction alternates between adjacent sites, indicating that the stripe CDW is an antiferroelectric phase. We also calculated a uniform-phase CDW, where the charge transfer is unidirectional and produces a net polarization, making it a ferroelectric CDW phase (a more detailed discussion of this phase and its competition with the antiferroelectric CDW can be found in the SI Section C).

\begin{figure}[t!]
    \centering
    \includegraphics[width =\columnwidth]{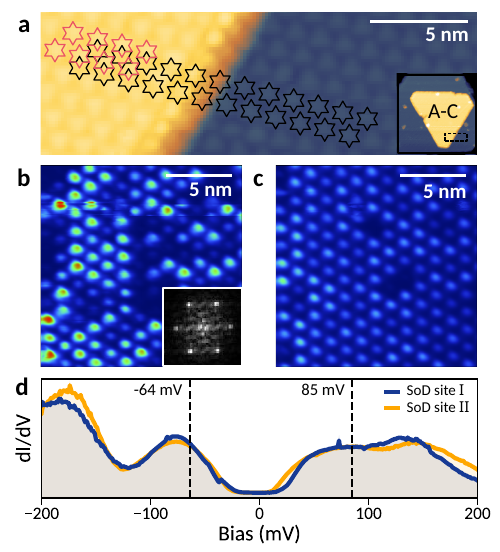}
    \caption{ A-C stacking 1T-TaSe$_2$ as a band insulator. \textbf{a,} A terrace of a typical bilayer island of A-C stacking. $V = 1$ V, $I = 10$ pA. The SoD cluster centres in the upper layer and bottom layer are marked with differently shaded stars. \textbf{b,} \textbf{c,} Constant-height DOS mapping at the energy of $V = -64$ mV and $V = 85$ mV. $I = 50$ pA. $V_{m} = 5$ mV. Inset: corresponding FT image showing CDW modulation. \textbf{d,} Spectra collected on two adjacent SoD centers (SOD site I and II) of the A-C bilayer. Both show a gap size of around 40 mV. $I = 300$ pA. $V_{m} = 5$ mV. }
    \label{fig:fig3} 
\end{figure}

\subsection*{Band insulator in the A-C 1T-TaSe$_2$ bilayer}

We now focus on our experimental results, starting with A-C-stacked bilayers. Figure \ref{fig:fig3}a shows an edge of an A-C bilayer island (inset). We mark the SoD centers on both layers with an extension across the terrace to show the overlay of the centers of CDW units, confirming the A-C-stacking of the layers. Figures \ref{fig:fig3}b and \ref{fig:fig3}c are constant-height LDOS maps taken at -64 mV and 85 mV, respectively. The inset FT image shows the spots corresponding to the SoD CDW without any additional spots. Figure \ref{fig:fig3}d shows the tunneling spectra at two neighboring CDW sites in the A-C bilayer. There are band onsets at $\pm \sim20$ mV with a zero-conductance hard gap at the Fermi level, which is consistent with the hybridization gap in the simulations. The non-modulated LDOS over neighboring CDW sites is also consistent with the band insulating picture from our theoretical results. Since A-A and A-C stacking are the main stacking ways of T-phase TaX$_2$ (X=S, Se), \cite{butler2020mottness,Lee2023,Wang2020}, our results also indicate that bilayer and bulk 1T-TaSe$_2$ are band insulators.

\begin{figure}[t!]
    \centering
    \includegraphics[width = \columnwidth]{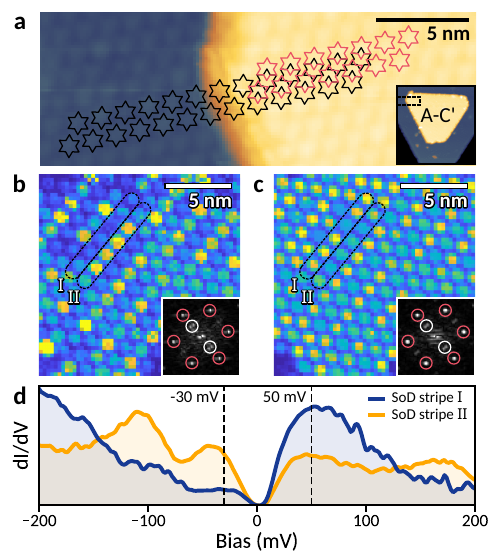}
    \caption {Correlation-driven stripe CDW phase in A-C$'$ bilayer. \textbf{a,} A terrace of a typical bilayer island of A-C$'$ stacking. $V = 1$ V, $I = 10$ pA. \textbf{b,} \textbf{c,} \dIdV maps collected from grid measurement showing the stripe CDW phase. The neighboring CDW rows alternate in intensity (repeating stripes labeled I-II) and form a double periodicity modulation. Insets: corresponding FT image. Peaks in the red circles correspond to the CDW sites, and peaks in the white circles correspond to the new striped phase. $V = -30$ mV (panel b). $V = 50$ mV (panel c). $I = 20$ pA. \textbf{d,} Spectra collected on SoD sites belonging to stripes I and II. $I = 300$ pA. $V_{m} = 1$ mV.}
    \label{fig:fig4}
\end{figure}

\subsection*{Ferroelectric CDW in the A-C$'$ 1T-TaSe$_2$ bilayer}

Figure \ref{fig:fig4}a shows the edge of a bilayer with A-C$'$ stacking. We performed the stacking analysis the same way as in Figure \ref{fig:fig3}a, confirming A-C'-stacking in this case. The stacking difference gives rise to strikingly different LDOS as illustrated in Fig.~\ref{fig:fig4}b and c. Figure \ref{fig:fig4}b is a \dIdV map collected from a grid measurement at -30 mV, where a stripe CDW pattern---bright and dark CDW rows---is resolved. The corresponding FT image (inset of Fig.~\ref{fig:fig4}b) shows a pair of FT peaks (in white circles) with double periodicity of the original CDW. The direction of the stripes is along the dimer chain formed between the CDWs of the top and bottom layers. This agrees with the theoretically suggested mechanism where the striped phase arises due to the charge transfer between the top and bottom layers. Figure \ref{fig:fig4}c presents the \dIdV map measured in the same region at 50 mV, where the contrast of the CDW rows is reversed compared to that observed at negative bias. Spectroscopy results over several neighboring rows are shown in the SI Section B. 
\dIdV spectra from neighboring CDW units (bright and dark) are shown in Figure \ref{fig:fig4}d. There is a small gap with band onsets at $\pm \sim8$ mV. The bright and dark CDW units have opposite spectral weight at positive and negative bias, in agreement with the charge transfer mechanism suggested theoretically. 

\begin{figure}[t!]
    \centering
    \includegraphics[width = \columnwidth]{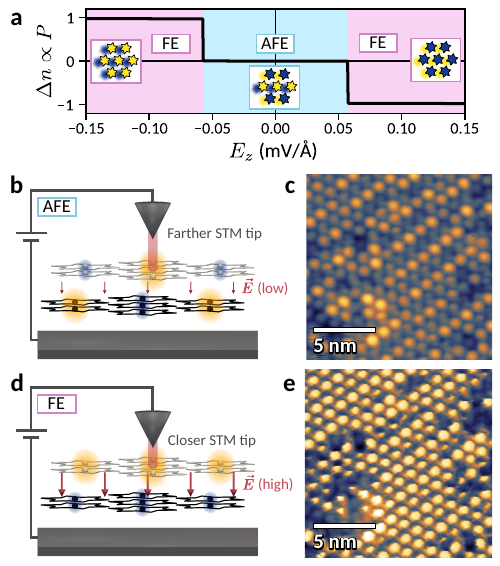}
    \caption{\textbf{a} Difference in occupation of upper and lower A-C$'$ layers as a function of electric field strength, showing switching from an antiferroelectric at low electric field to ferroelectric phases with opposite polarization $P$. \textbf{b,c} Schematic (b) and STM topography (c) of the striped antiferroelectric phase when the tip field is low. $V = -30$ mV. $I = 5$ pA. \textbf{d,e} Schematic (d) and STM topography (e) of the uniform ferroelectric phase when the tip field is high. $V = -30$ mV. $I = 1$ nA.}
    \label{fig:fig5}  
\end{figure}

Finally, we analyze the electrical tunability of the stripe CDW phase displayed by the A-C$'$ bilayer.
The TB model previously described, in the experimental regime, predicts that two competing phases with similar energy can be stabilized in the A-C$'$ bilayer. The striped antiferroelectric phase with alternating charge modulation between the top and the bottom 1T-TaSe$_2$ layer, observed experimentally (Fig.~\ref{fig:fig4}), is narrowly favored by the model. Its total energy is lower than that of a uniform ferroelectric phase in which there is a constant charge difference between the top and the bottom layers (illustrated in Fig.~\ref{fig:fig5}a). However, a perpendicular electric field $\vec{E} = (0,0, E_{z})$ between the layers can drive the A-C$'$ bilayer from the striped antiferroelectric phase to the uniform ferroelectric phase, by promoting charge transfer to one of the layers (Fig.~\ref{fig:fig5}a). The total polarization $P$ in the $z$-direction is proportional to the occupation difference of the upper and lower layers $\Delta n$ as a function of the external electric field. We find computationally that switching occurs at $E_{z} = 0.058$ mV/Å from the antiferroelectric phase to the ferroelectric phase. Tuning the tip field from low to high induces the switch from the antiferroelectric striped phase (Fig.~\ref{fig:fig5}b) to the ferroelectric uniform phase (Fig.~\ref{fig:fig5}d). We can test this by carrying out experiments at different tip-sample distances and hence, electric fields. This can be controlled by the tunneling current setpoint as illustrated in Figs.~\ref{fig:fig5}b-e. At a current setpoint of 5 pA (Fig.~\ref{fig:fig5}c), the system stays in the striped antiferroelectric phase. Increasing the setpoint up to 1 nA (Fig.~\ref{fig:fig5}e), the contrast of stripes disappears, and the system switches to a uniform ferroelectric state.

\section*{Conclusions}

In summary, we have demonstrated that bilayer 1T-TaSe$_2$ provides a platform where interlayer stacking engineering enables the emergence of distinct correlated electronic phases driven by the competition between interlayer hybridization and Coulomb interactions. Using STM/STS measurements combined with \emph{ab initio} calculations and low-energy modeling, we identified two stacking geometries, A-C and A-C$'$, which realize quasi-1D interacting fermionic chains with fundamentally different electronic ground states. While the A-C stacking is governed by interlayer dimerization and develops a hybridization band-gap insulating phase, the A-C$'$ stacking is dominated by interlayer Coulomb interactions that stabilize a spontaneous charge imbalance between layers, giving rise to an out-of-plane ferroelectric alternating polarization.

Our results establish an unconventional mechanism for ferroelectricity in van der Waals materials, where the polar order emerges primarily from strong electronic correlations in a flat-band system rather than from structural inversion symmetry breaking associated with lattice distortions or sliding reconstruction. Furthermore, the stabilization and electrical switching of ferroelectric and antiferroelectric interchain configurations demonstrate the tunability of the correlated polar state and highlight the rich interplay between charge order, dimensionality, and electronic interactions in coupled Star-of-David lattices.
More broadly, our work extends the phase diagram of layered 1T TMDs with a ferroelectric CDW phase driven by electronic interactions, opening new opportunities for realizing electrically controllable correlated states in flat-band quantum materials. These results establish stacking-engineered SoD heterostructures as a promising platform to explore emergent ferroic orders and correlation-driven functionalities.

\section*{Methods}

\subsection*{Sample preparation} 

1T-TaSe$_2$ was grown by molecular beam epitaxy (MBE) on highly oriented pyrolytic graphite (HOPG) under ultra-high vacuum conditions (UHV, base pressure $\sim1\times10^{-10}$ mbar). HOPG crystal was cleaved and subsequently out-gassed at $\sim600^\circ$C in UHV. High-purity Ta was evaporated from an electron-beam evaporator. Se was evaporated from a Knudsen cell using Se powder ($99.9\%$, Merck). Before growth, the flux of Ta was calibrated on an Au(111) at $\sim1$ monolayer per hour. The sample was grown at a Se background pressure of $\sim1\times10^{-8}$ mbar for 30 minutes. Before the growth, the HOPG substrate temperature was stabilized at $\sim550^\circ$C. 

\subsection*{STM measurements}

After the preparation, the sample was inserted into the low-temperature STM (Createc LT-STM) connected to the same UHV system, and subsequent experiments were performed at $T = 5$ K. STM images were taken in the constant-current mode. d$I$/d$V$ spectra were recorded by standard lock-in detection while sweeping the sample bias in an open feedback loop configuration, with a peak-to-peak bias modulation specified for each measurement and at a frequency of 757 Hz.

\subsection*{\textit{Ab initio} calculations} 

We have performed \textit{ab initio} electronic structure calculations based on density functional theory\cite{HK,KS} (DFT) in A-C and A-C' bilayers of TaSe$_2$. Calculations were carried out with the all-electron full-potential linearized augmented-plane-wave method as implemented in Elk \cite{elk}. We have used the generalized gradient approximation in the Perdew-Burke-Ernzerhof scheme (GGA-PBE) for the exchange-correlation functional \cite{PhysRevLett.77.3865}. 
The atomic positions of both A-C and A-C' stacking configurations were relaxed before analyzing their electronic structure. The results presented are converged with respect to all the parameters, considering a $6\times 6 \times 1$ k-mesh, and a vacuum spacing of 30 \AA.

\subsection*{Mean-field calculations with the low-energy model} 

The minimal TB model for quasi-1D interacting chains used to describe the bilayer configurations takes the form:

\begin{equation}\label{eq_H}
H = H_0 + H_U + H_{V},
\end{equation}

where $H_0$ is the single-particle hopping term:

\begin{equation}\label{eq_H0}
H_0 = \sum_{ij \in l, \,s}
t_{ij} \, c^\dagger_{is} c_{js}
+ 
\sum_{ij \notin l, \,s}
t_{\perp, ij} \, c^\dagger_{is} c_{js},
\end{equation}

with $s=\uparrow, \downarrow$, and $ij \in l$ and $ij \not\in l$ denoting sites belonging to the same layer and opposite layers, respectively. The intralayer hopping between nearest neighbors $t = 3$ meV is much smaller than the hopping between interlayer first neighbors $t_{\perp} + \delta t_{\perp}$ and second neighbors $t_{\perp} - \delta t_{\perp}$, as fitted to the \textit{ab initio} calculations from Fig.~\ref{fig:fig2}a-b.

The interacting part consists on the on-site Coulomb interaction $U$:

\begin{equation}
H_U = 
\sum_i U \, c^\dagger_{i\uparrow} c_{i\uparrow} c^\dagger_{i\downarrow} c_{i\downarrow}, 
\end{equation}

as well as interactions $V$ between the closest neighboring sites: the interlayer first- and second-nearest neighbors.

\begin{equation}
H_{V} = \sum_{\substack{ij \notin l, \\s,s'}} V_{{ij}} \, c^\dagger_{is} c_{is} c^\dagger_{js'} c_{js'}.
\end{equation}

The full model is solved by performing a mean-field approximation, making the replacement
\begin{align*}
    c^\dagger_{is} c_{is} c^\dagger_{js'} c_{js'} \approx& \; \langle c^\dagger_{is} c_{is} \rangle c^\dagger_{js'} c_{js'}  + \langle c^\dagger_{js'} c_{js'} \rangle c^\dagger_{is} c_{is}\\
    &- \langle c^\dagger_{js'} c_{is} \rangle c^\dagger_{is} c_{js'} - \langle c^\dagger_{is} c_{js'} \rangle c^\dagger_{js'} c_{is}\\
    &- \langle c^\dagger_{is} c_{is} \rangle \langle c^\dagger_{js'} c_{js'} \rangle + \langle c^\dagger_{js'} c_{is} \rangle \langle c^\dagger_{is} c_{js'} \rangle
\end{align*}
in the interaction terms. The resulting mean-field equations are solved self-consistently to find the variational single-particle states $\psi^{\dagger}_n$ and build the variational Hartree-Fock wavefunction $| \Omega_{\text{MF}}\rangle = \prod_n \psi^\dagger_n |0 \rangle$.\cite{pyqula}
The solution of the mean-field Hamiltonian features a complex phase diagram that we discuss in the main text. In particular, it leads to interaction-driven ferroelectric and antiferroelectric CDW phases.

\subsection*{Modeling of the external electric field in the A-C' bilayer}

The addition of an interlayer bias $\Delta = E_{z}\,d_{z}$, where $d_{z} = 6.29$ Å is the interlayer distance, can be simulated by including an extra term $H_{\Delta}$ in the tight-binding model:

\begin{equation}
H_\Delta = \frac{\Delta}{2} \sum_{i,s} \chi_i c^\dagger_{is} c_{is},
\end{equation}

where $\chi_{i} = +1$ if site $i$ belongs to the upper layer and $\chi_{i} = -1$ if it belongs to the lower layer.

\section*{Data availability}
All the data supporting the findings are available from the corresponding authors upon request.

\section*{Acknowledgements}
This research made use of the Aalto Nanomicroscopy Center (Aalto NMC) facilities and was supported by the Research Council of Finland Projects Nos.~371757, 369367, and 347266, EU Horizon Europe Marie Skłodowska-Curie Actions 101154353, ERC AdG GETREAL (no.~101142364), and ERC CoG ULTRATWISTROICS (no.~101170477). We acknowledge the financial support of the Finnish Ministry of Education and Culture through the Quantum Doctoral Education Pilot Program (QDOC VN/3137/2024-OKM-4), the Research Council of Finland through the Finnish Quantum Flagship project (358877, Aalto University) and the Finnish Centre of Excellence in Quantum Materials QMAT (No. 374166), and the computational resources provided by the Aalto Science-IT project.

\section*{Contributions}

Z.W., and P.L. initiated and conceived this project. Z.W., M.A., Y.D., and R.D. carried out the STM/STS measurements. Z.W. performed the sample growth. A.V.M., J.L.L., and A.O.F. performed the theoretical modeling. Z.W., A.V.M., and A.O.F. analyzed the data and wrote the manuscript with feedback from all the co-authors.

\section*{Competing interests}
The authors declare no competing interests.

\bibliography{cdw}

\end{document}